\documentclass[prl,aps,twocolumn,preprintnumbers,amsmath,amssymb,showpacs]{revtex4}
\usepackage{graphicx}\usepackage{epsfig}
\begin{document}
\title{
 Expansion dynamics of Lennard-Jones systems}
%
%
\author{M. J. Ison$^{1,2}$\footnote{Present address: Department of Engineering, 
University of Leicester, LE1 7RH, UK}} 
\author{F. Gulminelli$^{2}$\footnote{member of the Institut Universitaire de France}}
\author{C. Dorso$^{1}$}
\affiliation{$^{1}$~Departamento de F\'{\i}sica, Facultad de Ciencias Exactas y Naturales,
Universidad de Buenos Aires, Buenos Aires (1428), Argentina}  
\affiliation{$^{2}$~LPC Caen, ENSICAEN, Universit\'e de Caen, CNRS/IN2P3, Caen, France}
%
%
\date{\today}
%
%
\begin{abstract}
The dynamics of the expansion of a Lennard-Jones system, initially confined 
at high density and subsequently expanding freely in the vacuum, is 
confronted to an expanding statistical ensemble, derived in the diluted 
quasi-ideal Boltzmann approximation. The description proves to be fairly 
accurate at predicting average one-body global observables, but important 
deviations are observed in the configuration-space structure of the events.
Possible implications for finite expanding physical systems are outlined.
\end{abstract}
%
%
%

\maketitle
%
%

\section{Introduction}

When a finite isolated unbound system decays in the vacuum, its decay pattern
is often characterized by an ordered kinetic component: the collective flow.
This is the case in clusters dissociation induced by photo-ionization
 \cite{Catherine,Haberland,Brockhaus,Martinet} or charge transfer 
collisions \cite{Farizon,Gluch}. Here, the condensed matter bulk limit
encourages, in principle, an interpretation in terms of liquid-vapor transition.
 However, the obvious fact that all vapor flows out and no vapor comes back in, 
makes it difficult to push the analogy further, and
the dynamical evaporation can only be interpreted thermodynamically
 \cite{Catherine,Haberland,Farizon} by making use of a time-dependent temperature 
within the concept of an evaporative ensemble \cite{Klots,Calvo}.
Flow is also a basic feature of heavy ion collisions, where the products 
of fragmentation reactions show a velocity preferentially oriented in the radial
direction \cite{Wci}. If in the Fermi energy regime 
and in the associated multi-fragmentation phase transition these collective flows may be 
only a perturbation in the global energetics, this is not true at SIS energies 
(between 0.2 and 2.0 GeV/nucleon), where 
they are likely to influence light cluster formation by coalescence \cite{Reisdorf}. 
In the ultra-relativistic regime the ordered and disordered motions become comparable in magnitude
 \cite{shm}, and collective flows are believed to play an essential role in 
the characteristics of the transition to the quark-gluon plasma observed in the 
Relativistic Heavy Ion Collider data \cite{Shuryak,Ko,hydro}. 

In the initial stage of a nuclear collision the complexity of the dynamics is such 
that a statistical analysis of the system might prove useful, even at energies as low 
as in the Fermi regime.
If this equilibrium stage occurs at high density \cite{Campi}, the final-state interaction may still
be important in the subsequent evolution, and it is not clear how the final partitions
at the freeze-out stage will be modified.
In a previous paper \cite{Ariel} we have addressed this issue at a classical level 
through molecular dynamics simulations of a schematic Lennard-Jones system \cite{Dorso}, 
initially thermalized at high (supercritical) density,
and subsequently freely expanding in the vacuum. We have shown that the dynamics of the 
expansion leads to a considerable increase of fluctuations. By the time when partitions are
settled and the formed pre-fragments cease to interact (freeze-out), these fluctuations are
qualitatively similar to the ones expected from a thermal system at reduced subcritical
density. A naturally arising question is then whether the expanded system at freeze-out can still
be treated as a statistical equilibrium at a lower density.

Our numerical experiments can only address in a very partial and incomplete way
this important question, which does probably not have a unique answer. 
In the evaporation regime accessed in cluster experiments the above-mentioned freeze-out 
configuration does not exist, and a statistical treatment equivalent to the dynamical process
has to be adapted to the time window of the experiment \cite{Catherine}.
Conversely in a heavy ion collision, time scales are so short that well defined 
freeze-out time(s) can be identified. However, such collisions may correspond
to a dynamical process very different from the free expansion of a dense system of classical
particles. In particular, it is important to stress that macroscopic statistical models
in all energy regimes \cite{Bondorf,shm} suppose that the statistical hypothesis applies
at the freeze-out time, and in this sense they by-pass by construction the problem
of the out-of-equilibrium evolution up to freeze-out.

In different physical situations flow appears to settle early in the dynamics.
Notably, this is the case of central nuclear collisions, where flow is associated to an 
initial compression of the dinuclear system. In such a situation equilibrium, if ever reached, 
is approached when the system is still strongly interacting, 
and we may expect the successive dynamics to deeply modify the system configurations. 
In these cases, the model of an initially equilibrated dense molecular system, which then 
is freely expanding in the vacuum, may bear some pertinent information.

%
%
In this paper we compare freeze-out configurations 
of a freely expanding Lennard-Jones system with different equilibrium models
for the same system. We show that, if statistical models are reasonably 
correct as far as average quantities are concerned, important differences 
can be seen in the fragmentation patterns. In particular, close to the liquid-gas
phase transition unstable configurations, inaccessible to equilibrium
models, appear to dominate the dynamics of the expansion.

%

\section{ The isobar microcanonical ensemble}

The system under study is composed of excited drops
made up of $N=147$ particles interacting via a Lennard-Jones (LJ) $6-12$ potential $v_{LJ}(r)$ with a cutoff 
radius $r_c=3\sigma$. Energies are measured in units of the potential well ($\epsilon$), 
$\sigma$ characterizes the radius of a particle and $m$ is its mass. We adopt adimensional
units for energy, length, and time such that $\epsilon=\sigma=1$, $t_0=\sqrt{\sigma^2m/48\epsilon}$.   
The initial condition is given by the microcanonical isobar statistical
ensemble described by the probability for each microstate $(n)$
\begin{equation}
p^{(n)}_0 
= \frac{\delta \left ( E - {H_{LJ}^{(n)}}\right )}{W_{\lambda}\left ( E \right )}
\exp \Bigl [-\lambda \sum_{i=1}^N (r_{i}^{(n)})^2\Bigr ] 
,  \label{t0}
\end{equation}
where $r_{i}^{(n)}$ is the position of particle $i$ within the microstate $(n)$, 
${H_{LJ}^{(n)}}$ is the corresponding Lennard-Jones energy, 
$\lambda$ is a Lagrange multiplier constraining a finite size, and
\begin{equation}
W_{\lambda}\left ( E \right )=\sum_{(n)} \exp \Bigl [-\lambda \sum_{i=1}^N (r_{i}^{(n)})^2\Bigr ] 
\delta \left ( E - H_{LJ}^{(n)}\right ) \label{entropy}
\end{equation}
is the associated density of states or partition sum.
The distribution eq.(\ref{t0}) is the minimum biased probability distribution
for an isolated finite system with a finite size measured by its mean square
radius \cite{annals}
\begin{equation}
\langle R^{2}\rangle = \langle \sum_{i=1}^N (\vec{r}_{i}^{(n)})^2 \rangle_n ,
\end{equation}
where the average is taken over microstates.
To generate the statistical ensemble eq.(\ref{t0}), we numerically proceed as follows.
A harmonic potential with spring constant
$k=m \omega^2$, with $\omega=0.1 t_0^{-1}$ is added to the Hamiltonian, 
and the system is coupled to a thermostat using the Andersen technique \cite{Andersen} 
to achieve equilibrium inside the oscillator. 
In brief, this is attained by stochastic impulsive forces that act
occasionally on randomly selected particles. After each collision the
particle is endowed with a new velocity drawn from a
Maxwell-Boltzmann distribution at the desired temperature $\beta^{-1}$. 
Between stochastic collisions, the system evolves at constant energy. It has been proved that, 
under some general conditions \cite{Andersen}, the constant energy shells are visited 
according to their Boltzmann weights, which in turn implies that the ensemble of 
configurations at different times constitutes a canonical ensemble at the thermostat 
temperature, distributed as
\begin{eqnarray}
p^{(n)}_{cano}
&=& \frac{1}{Z_{\beta,k}}
\exp -\beta \sum_{i=1}^N \Bigl [ 
\frac{(\vec{p}_{i}^{(n)})^2}{2m} +\frac{k}{2} (\vec{r}_{i}^{(n)})^2  \Bigr ]+ \nonumber \\  
&& - \beta \sum_{i < j}^N  v_{LJ}\left ( |\vec{r}_{i}^{(n)}-\vec{r}_{j}^{(n)}|\right )
\label{t0cano}
\end{eqnarray}
Microcanonical ensembles are extracted at different energies 
by sorting the events of the canonical distributions according to their energy,
excluding the contribution of the confining potential.
These dense configurations correspond for all energies to the supercritical part of
the Lennard-Jones phase diagram \cite{Dorso_PD}.

\section{Comparison with the free expanding system}

\subsection{Choice of the comparison time}

The equilibrated configurations are let evolve in the vacuum for a time long enough
that the chemical composition of the system is settled (freeze-out time) 
\footnote{for technical details on the implementation see ref. \cite{Ariel}.}.
 An interesting observable to study the freeze-out properties of the system 
is given by the normalized kinetic energy fluctuation 
$A_K=N\sigma^2_K/\langle K \rangle ^2$. This quantity shows a saturation 
when the sharing between kinetic and potential energy finishes, which
physically corresponds to the formation of surfaces within the system, which in turn
implies a chemical stabilization of the cluster properties \cite{Ariel}.
The behavior of this fluctuation with time is displayed for four representative
energy states in the upper part of Figure 1. 

%
%
\begin{figure}[b!]
\setlength{\abovecaptionskip}{-10pt}
\begin{center}
\includegraphics[angle=-90, width=.97\columnwidth, trim=1cm 0cm 0.5cm 0cm,clip=true]{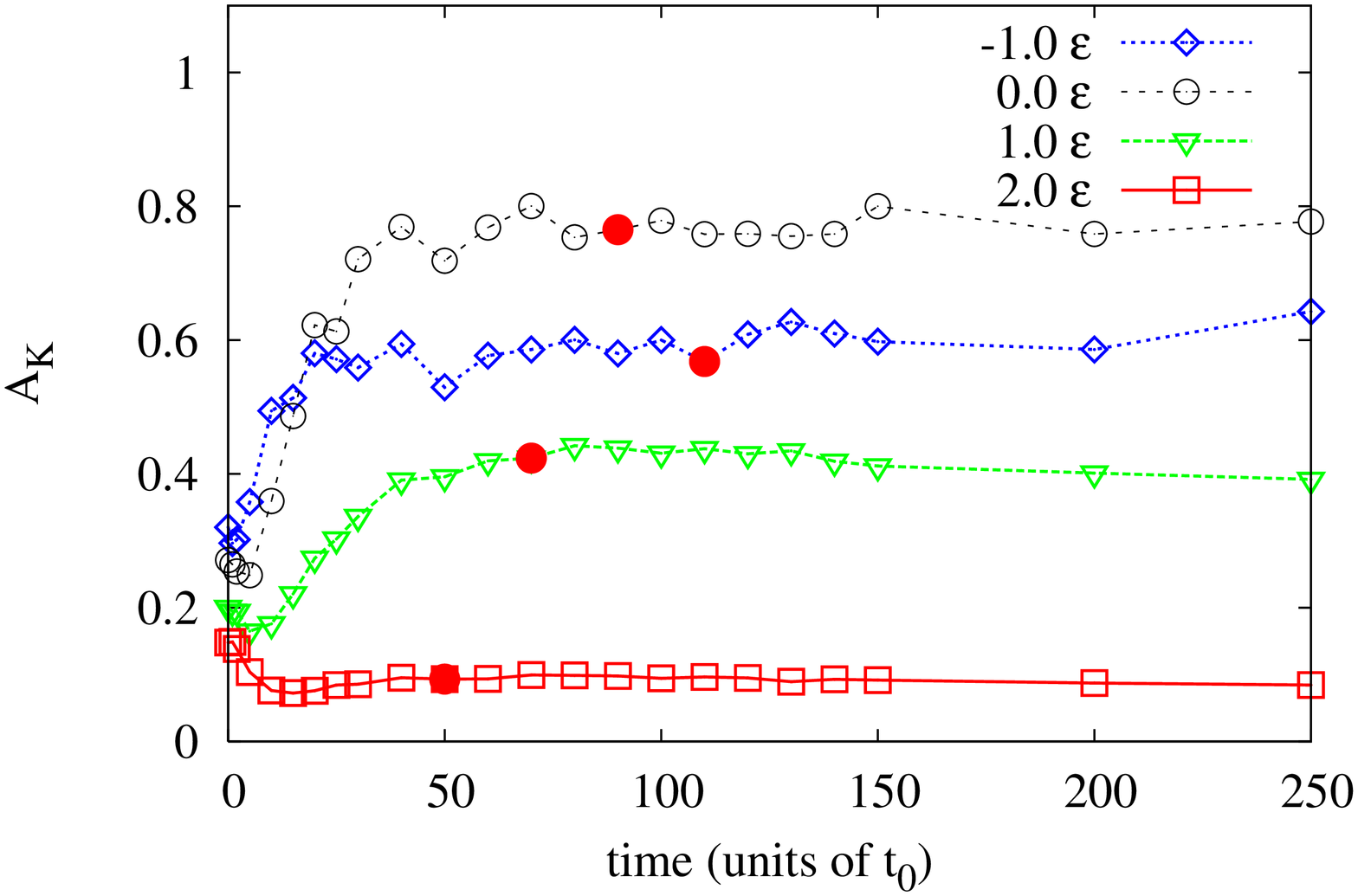}
\includegraphics[angle=-90, width=.94\columnwidth, trim=1.2cm 0cm 0cm 0cm,clip=true]{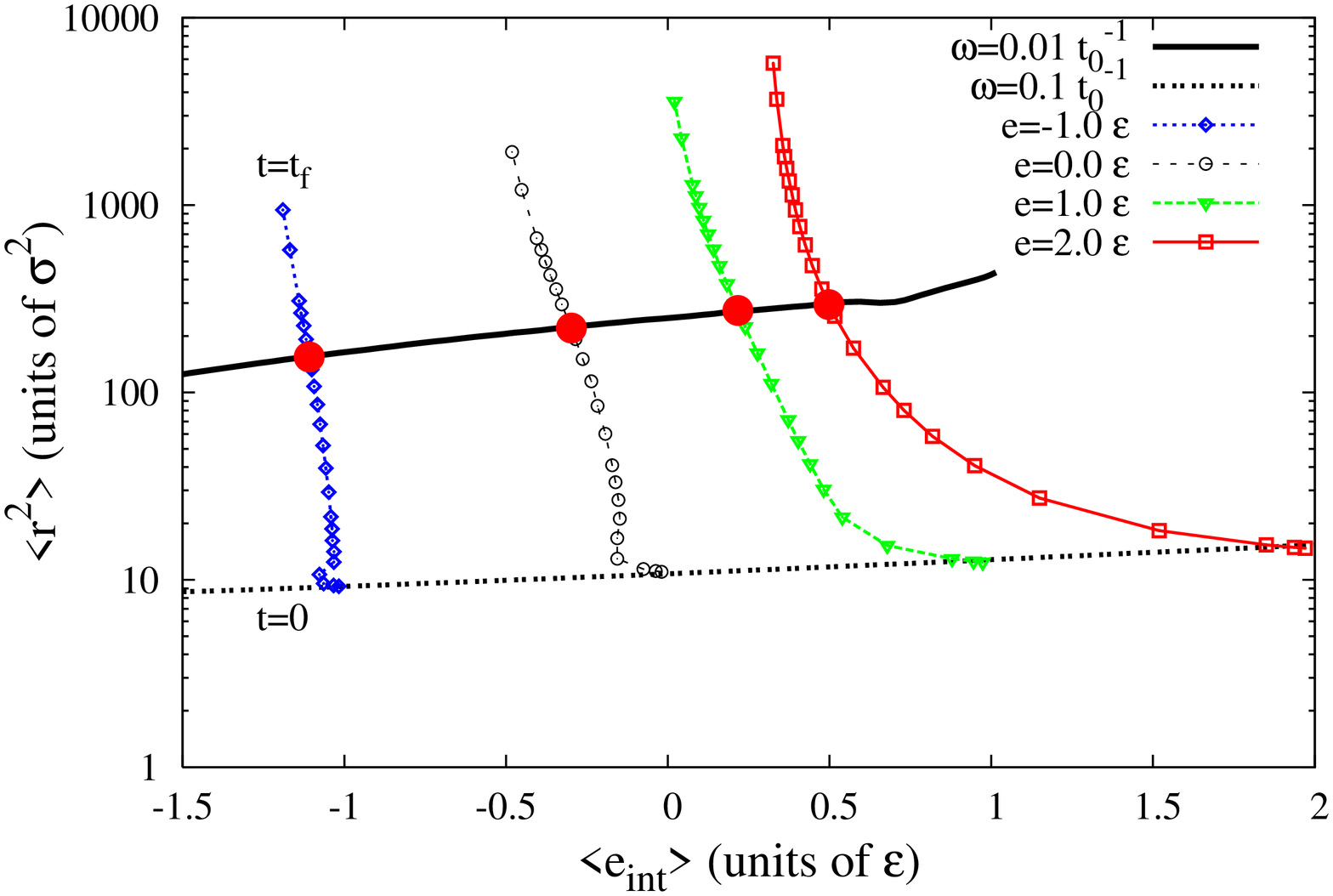}
\end{center}
 \caption{\label{fig:fig1}
(Color online) Lower part: mean square radius as a function of the internal energy of a 
confined $LJ$ system in a harmonic
trap with $\omega=0.01t_0^{-1}$ (full line) and $\omega=0.1t_0^{-1}$ (dashed line), 
compared to 
$LJ$ systems in free expansion, at different total energies and times between $0t_0$ and $t_f=250t_0$(symbols). 
Upper part: kinetic energy fluctuation $A_K$ (see text) as a function of time for the free expanding systems at 4 different energies.  
The large circles in both pictures correspond to the time at which the confined and the free 
expanding system have the same mean spatial extension. All quantities are in expressed reduced units.
}
\end{figure}  

During the time evolution, the initially disordered kinetic energy is partially 
converted into collective motion, defined as
\begin{equation}
E_{flow}^{(n)}(t)= N \frac{(\vec{p}_{r}^{(n)})^2}{2m}=\sum_{i=1}^N \frac{1}{2m}
\left ( \vec{p}_{i}^{(n)}\cdot \frac{\vec{r}_{i}^{(n)}}{{r}_{i}^{(n)}}\right )^2.
\label{K_flow}
\end{equation}
The internal energy at each time is correspondingly given by 
$E_{int}=K_{int}+V_{LJ}$, where 
\begin{equation}
K_{int}^{(n)}(t)=\sum_{i=1}^N \frac{1}{2m}
\left ( \vec{p}_{i}^{(n)}- \vec{p}_{r}^{(n)}\right )^2,
\label{K_int}
\end{equation}
and $V_{LJ}^{(n)}=\sum_{i\neq j}  v_{LJ}\left ( |\vec{r}_{i}^{(n)}-\vec{r}_{j}^{(n)}|\right ) $ 
is the interaction part of the Lennard-Jones energy.

The time evolution of the system with different initial energies is represented 
in the $\langle E_{int}\rangle$ versus $\langle R^2\rangle$ plane in the lower part of Figure 1.
We can see that the time evolution corresponds to a rapid expansion
and a development of collective flow which enlarges with increasing total energy.
If at the entropy-saturation time or freeze-out time $t_f$ the system were still
close to a statistical equilibrium, it should be described by eq.(\ref{t0}) 
with $E=\langle E_{int}\rangle (t_f)$ and a new volume constraint $\omega$
such that the average over the statistical ensemble (\ref{t0}) of the 
mean square radius is $\langle R^2 \rangle =\langle R^2\rangle (t_f)$.
The mean square radius of an equilibrated system with the looser constraint
$\omega=0.01 t_0^{-1}$ is also represented in the lower part of Figure 1.
The points where this curve crosses the time evolution of the freely expanding 
systems give the times we have chosen to compare the statistical and dynamical
ensembles. As it can be observed from the upper part of the figure, 
for all chosen energies these times are long enough for the configurations
to be safely considered as frozen. At later times the dynamics will not further affect 
energy and fragment partitions, meaning that the comparison we show will be 
pertinent also for later times. 

\subsection{Energy sharing and effective temperature}

The time dependent  sharing of the internal energy $E_{int}=K_{int}+V$ between 
the kinetic and interaction component is shown in Figure 2, which displays the average 
disordered kinetic energy $\langle K_{int}\rangle$ as a function of the internal energy
for the different evolutions. The equilibrium correlations for the initial and final
state are also represented. 
%
%
\begin{figure}[htbp]
\setlength{\abovecaptionskip}{-10pt}
\begin{center}
\includegraphics[angle=-90, width=.94\columnwidth, trim=1.2cm 0cm 0cm 0cm,clip=true]{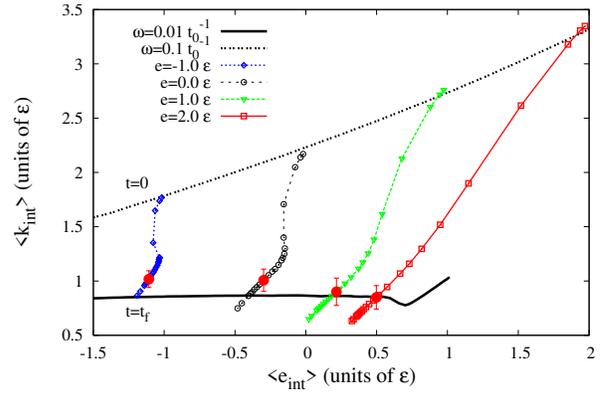}
\end{center}
\caption{\label{fig:fig2}
(Color online) Average kinetic energy per particle as a function of the internal energy per particle 
of a confined $LJ$ system in a harmonic trap with $\omega=0.01t_0^{-1}$ (full line) and 
$\omega=0.1t_0^{-1}$ (dashed line) 
compared to 
$LJ$ systems in free expansion at different total energies at times between $0t_0$ and 
$t_f=250t_0$ (symbols). Emphasized symbols correspond to the time at which the confined and the free 
expanding system have the same mean spatial extension.
 }
\end{figure}   

We can see that the equilibrium sharing is well verified at the freeze-out time,
except a slight underestimation at the lower energies.
This result implies that the kinetic energy at freeze-out, once the collective component 
is subtracted, can be used as a thermometer measuring a physically well defined 
temperature for the expanding system.

The deviation at low energy is also interesting.
At $E/N=-1\epsilon$ the system evolution is essentially a simple evaporation of monomers 
and light fragments at the surface of a single excited condensed drop.
The poor reproduction by the statistical ansatz (\ref{t0}) of the drop kinetic energy
implies that the evaporation rate can only approximately be described in terms of an 
effective pressure \cite{moretto}, due to the irreducible time dependence of the 
evaporation process.

%
%
\begin{figure}[b!]
\setlength{\abovecaptionskip}{10pt}
\begin{center}
\includegraphics[angle=0, width=.9\columnwidth, trim=0cm 0cm 0cm 0cm]{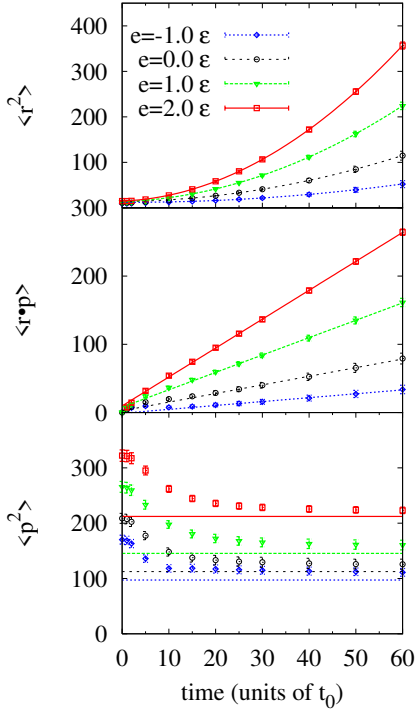}
\end{center}
\caption{\label{fig:fig3}
(Color online) Time dependence of the average square radius, radial momentum, 
and square momentum (all in reduced units) for the freely expanding system at four different energies. 
Symbols: numerical simulations, where the internal square momentum of the largest cluster 
recognized at asymptotic times is subtracted. Lines: free gas evolution 
eqs.(\ref{eqdyn1},\ref{eqdyn2},\ref{eqdyn3})
corresponding to the statistical ansatz eq.(\ref{breath}) with an initial flow. \\
}
\end{figure} 

\subsection{Statistical treatment of the expansion dynamics}

At first sight it may be surprising that the best adequacy to the equilibrium picture 
is obtained at the two highest energies, where  
the flow contribution is the most important, and the dynamics is the fastest (see Figure 1).
This counter-intuitive result can be understood if we consider that the structure
(\ref{t0}) for the microstate distribution is exactly preserved by the time evolution in the case of non-interacting
particles or local interactions, as we now show \cite{annals,matias}.
For a Hamiltonian system, the time dependence of any mean observation $\langle A\rangle$
is given by $\partial_t \langle A\rangle = - \langle \{ H, A\} \rangle$.
If the system is non-interacting ($H=K$) this immediately gives 
\begin{eqnarray}
\partial_{t} \langle \vec{r}^2 \rangle &=& \frac{2}{m} \langle \vec{{r}} \cdot \vec{{p}}\rangle 
\nonumber \\
\partial_{t} \langle \vec{{r}} \cdot \vec{{p}}\rangle &=& \frac{1}{m} \langle \vec{p}^2 \rangle 
\label{eqdyn} \\
\partial_{t} \langle \vec{p}^2 \rangle &=& 0
\nonumber
\end{eqnarray}
It is easy to show that these same relations hold for a non-Hamiltonian dynamics in the presence of a Boltzmann
collision integral accounting for local two body interactions.
Let us consider an initial condition given by the equilibrium distribution eq.(\ref{t0cano})
(with $v_{LJ}=0$), imposing an average kinetic energy and mean square radius through the Lagrange multipliers 
$\beta$ and $\lambda=\beta m \omega^2/2$ constraining respectively the observables $\langle \vec{p}^2 \rangle$
and $\langle \vec{r}^2 \rangle$. At any successive time the exact evolution of these
observables will be given by eqs.(\ref{eqdyn}).
The minimum biased distribution fulfilling these time dependent constraints
and additionally conserving the total energy, is given at any time $t$ by
\begin{eqnarray}
p^{(n)}\left ( t \right ) 
&=& \frac{\delta \left ( E - {H_{LJ}^{(n)}}\right )}{W_{\lambda}
\left ( E,t \right ) }
\exp \Bigl [ -\tilde{\beta}\left ( t \right )  \sum_{i=1}^N 
\frac{(\vec{p}_{i}^{(n)})^2}{2m} + \nonumber  \\ 
&-& \lambda  \sum_{i=1}^N (\vec{r}_{i}^{(n)})^2 
+\nu\left( t\right) \sum_{i=1}^N
 {\vec{p}}_{i}^{(k)}\cdot {\vec{r}}_{i}^{(n)} \Bigr ]
  ,  \label{cano}
\end{eqnarray}
with 
\begin{equation}
\tilde{\beta} \left( t\right) = \frac{2\lambda}{m}t^{2} \;\;\; , \;\;\; 
\nu   \left( t\right) = \frac{2\lambda}{m} t .
\label{beta-nu}
\end{equation}
Using the associated dynamical equation (Liouville equation for the ideal gas or Boltzmann 
equation for the collision problem), it is possible to show \cite{annals}
that indeed eq.(\ref{cano}) is the exact evolution of eq.(\ref{t0cano})
under the action of the ideal gas Hamiltonian.

Eq.(\ref{cano}) can be interpreted as a radially expanding ideal gas in local equilibrium
with a time dependent temperature and pressure. Indeed the distribution can be written
as
\begin{eqnarray}
p^{(n)}\left ( t \right ) 
&=& \frac{\delta \left ( E - {H_{LJ}^{(n)}}\right )}{W_{\lambda}
\left ( E,t \right ) } \\ \nonumber
&\exp& \Bigl [ 
- \tilde{\beta}\left ( t \right )  \sum_{i=1}^N 
\frac{\left ( \vec{p}_{i}^{(n)}- m \tilde{h}\left ( t \right ) \vec{r}_i^{(n)}\right )^2 }{2m} \Bigr ]    
   ,  \label{micro-expan}
\end{eqnarray}
where $\tilde{h}=1/t$ represents a Hubblian factor. 

In the absence of an attractive inter-particle interaction the whole
momentum distribution participates to the flow dynamics. As noticed above, 
in the LJ case a part of the initial kinetic energy is converted into internal
energy of the clusters. Because of the spherical symmetry of the problem 
$\langle \sum_{i=1}^N \vec{p_i} \rangle=N\langle\vec{p_r}\rangle$, 
the two contributions are decoupled 
$N\langle \vec{p}^2\rangle/2m = \langle K_{int} \rangle + \langle E_{flow}\rangle$  
(see eqs.(\ref{K_flow},\ref{K_int})), and we can expect eqs.(\ref{eqdyn})
to be still approximately fulfilled, provided
the internal contribution to the kinetic energy is subtracted.
The time evolution of the average square radius, collective momentum, and square 
momentum are represented for different total energies in Figure 3.
At the beginning of the evolution the strong interactions acting in these 
dense supercritical systems modify the dynamics with respect to the ideal gas 
or diluted Boltzmann ansatz. However, starting from time $t\approx20t_0$
we can see that the trend predicted by eqs.(\ref{eqdyn})
is well verified. At $t\approx20t_0$ the system is still dense and homogeneous 
in first approximation \cite{Ariel} (see fig.1). We can then make the assumption 
that at $t\approx 20t_0\equiv t_1$ the initial condition (\ref{t0}) still describes the observed distribution
with the extra constraint of a collective flow 
$\langle {\vec{p}}\cdot {\vec{r}}\rangle \left ( t_1 \right ) \neq 0$,
\begin{eqnarray}
p^{(n)}\left ( t_1 \right ) 
&=& \frac{\delta \left ( E - {H_{LJ}^{(n)}}\right )}{W_{\lambda_1} }
\exp \Bigl [ - {\lambda_1}  \sum_{i=1}^N (\vec{r}_{i}^{(n)})^2 + \nonumber \\ 
&+& \nu_1  \sum_{i=1}^N
 {\vec{p}}_{i}^{(n)}\cdot {\vec{r}}_{i}^{(n)} \Bigr ]
  .  \label{t1}
\end{eqnarray}
In the ideal or Boltzmann gas limit, the successive evolution of the distribution
(\ref{t1}) is given again by eq.(\ref{cano}), with a modification of the time dependent constraints to account for the initial flow.
For the purpose of getting analytical results it is easier to consider an initial
condition in the canonical ensemble:
\begin{eqnarray}
p^{(n)}_{cano}\left ( t_1 \right ) 
&=& \frac{1}{Z_{\beta_1,\lambda_1} }
\exp \Bigl [ 
-  \sum_{i=1}^N {\beta_1}(\vec{p}_{i}^{(n)})^2 
-             {\lambda_1}(\vec{r}_{i}^{(n)})^2 + \nonumber \\ 
&+& \nu_1  \sum_{i=1}^N
 {\vec{p}}_{i}^{(n)}\cdot {\vec{r}}_{i}^{(n)} \Bigr ]
  .  \label{t1_cano}
\end{eqnarray}
Then the time dependent partition sum can be factorized $Z_{\tilde{\beta_1},\tilde{\lambda_1},\tilde{\nu_1} }= z_{\tilde{\beta_1},\tilde{\lambda_1},\tilde{\nu_1} }^N$ with
\begin{eqnarray}
z_{\tilde{\beta},\tilde{\lambda},\tilde{\nu} }(t)&=& 
\frac{1}{h^3} \int d^3r \int d^3p
 \Bigl[ \exp \Bigl( -\tilde{\beta_1} \left( t\right)  
\frac{\vec{p}^{2}}{2m} -\tilde{\lambda_1}\left( t\right) 
\vec{r}^{2} \nonumber \\
&+& \tilde{\nu_1}\left( t\right)  {\vec{p}}\cdot 
{\vec{r}}  \Bigr) \Bigr] \\ \nonumber
&=& \frac{2\sqrt{\pi}m}{h^3} \left ( 2 \tilde{\beta_1} \tilde{\lambda_1} - 
\tilde{\nu_1}^2 m \right) ^{-3/2}
. \label{breath}
\end{eqnarray}
The time dependent Lagrange parameters are given by
\begin{eqnarray}
\tilde{\beta}_1 \left( t\right) &=& \beta_1 - 2\nu_1 \Delta t + \frac{2\lambda_1}{m} \Delta t^{2} \nonumber \\
\tilde{\lambda}_1\left ( t \right ) &=& \lambda_1  \nonumber \\
\tilde{\nu}_1 \left( t\right) &=& \nu_1 - \frac{2\lambda_1}{m} \Delta t,
\end{eqnarray}
and the time dependent equations of state give for any $\Delta t=t-t_1>0$
the predicted evolution of the average observables:
\begin{eqnarray}
\label{eqdyn1} &&\langle p^2 \rangle = -2 m \frac{\partial \log z_{\tilde{\beta_1}\tilde{\lambda_1} \tilde{\nu_1}}}{\partial \tilde{\beta_1}(t)}
= \frac{6 m {\lambda_1}}{2{\beta_1}{\lambda_1}-m{\nu_1}^2}   \\
&&\langle r^{2} \rangle = -\frac{\partial \log z_{\tilde{\beta_1}\tilde{\lambda_1} \tilde{\nu_1}}}{\partial \tilde{\lambda_1}(t)}
= \frac{3\beta_1 - 6 \nu_1 \Delta t + \frac{6\lambda_1}{m} \Delta t^2}
{2{\beta_1}{\lambda_1}-m{\nu_1}^2 }  \label{eqdyn2} \\
&&\langle \vec{p}\cdot \vec{r} \rangle = 
\frac{\partial \log z_{\tilde{\beta_1}\tilde{\lambda_1} \tilde{\nu_1}}}
{\partial \tilde{\nu_1}(t)}
= \frac{3 m {\nu_1} - 6 \lambda_1 \Delta t}
{2{\beta_1}{\lambda_1}-m{\nu_1}^2} \label{eqdyn3}
\end{eqnarray}

The good adequacy displayed in Figure 3 between the numerical evolution 
and the ideal gas dynamics eqs.(\ref{eqdyn1},\ref{eqdyn2},\ref{eqdyn3}) shows that, 
even in the diabatic limit, as far as average global 
observables are concerned, the expansion can be represented at any time as a
statistical  equilibrium in the local rest frame eq.(\ref{cano}).

%
%
\begin{figure}[b!]
\setlength{\abovecaptionskip}{-10pt}
\begin{center}
\includegraphics[angle=-90, width=.95\columnwidth, trim=0cm 0cm 0cm 0cm,clip=true]{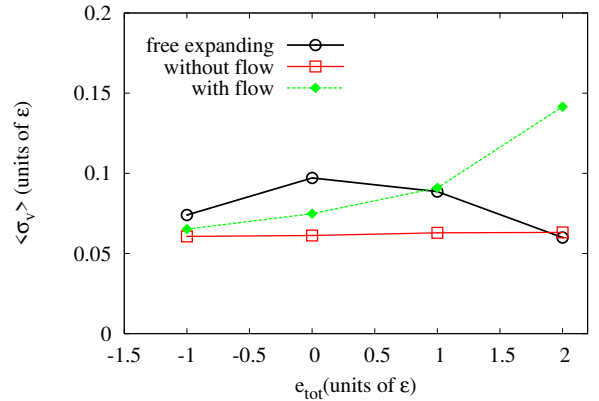}
\end{center}
\caption{\label{fig:fig4}
(Color online) Potential energy per particle fluctuations for the free expanding, constrained without 
flow (eq.(\ref{t0})), and constrained with flow (eq.(\ref{micro-expan})) systems. The abscissa
gives the total energy per particle of the free expanding system in all cases.}
\end{figure}   

\subsection{Deviations from equilibrium}

The good reproduction of time dependent average observables by the 
equilibrium ansatz does not mean that the configurations explored by the expanding system
coincide with the equilibrium configurations.
To look for deviations from equilibrium, we first show in Fig.4 the behavior of potential 
energy fluctuations. In the low energy liquid regime, as well as close to the liquid-gas
transition, the freeze-out fluctuations are very close to the prediction of the equilibrium
model. For the intermediate energies ($E/N=0\epsilon,E/N=1\epsilon$) however, these fluctuations are
underestimated by the equilibrium calculation. 

This underestimation can be partly due
to the energy conservation constraint. Indeed in the expansion of 
an isolated system the total energy is a constant of motion. This means that the potential 
energy can be converted both into internal kinetic energy, and into collective flow, 
inducing event-by-event fluctuations in both components. This effect is not considered 
in eq.(\ref{t0}) where the flow is absent, and the conservation law therefore
applies on the total internal energy. In other words, the equilibrium ansatz (\ref{t0})
misses flow energy fluctuations.

To quantify this effect and explore if it can explain the fluctuation underestimation, 
we have plotted in Figure 4 also the prediction of 
eq.(\ref{micro-expan}). Three models are then compared in this figure:
the free expansion at time $t_f$, and the equilibrium model in the 
local rest frame including (eq.(\ref{micro-expan})) or not (eq.(\ref{t0}))
flow fluctuations.
The three considered models are tuned to 
have in average the same internal energy $\langle E_{int}\rangle$ 
and spatial extension $\langle R^2 \rangle$. This internal energy
is fixed in each event $(n)$, $K^{(n)}+V^{(n)}=E_{int}=constant$, in the case of eq.(\ref{t0}); 
while it can fluctuate in the free expansion and also for eq.(\ref{micro-expan}), 
where $K^{(n)}+V^{(n)}=E_{int}^{(n)}+E_{flow}^{(n)}=E_{tot}=constant$.
We can see from Figure 4 that indeed flow fluctuations can enhance the potential energy 
variance, but they cannot explain the deviation as they do not exhibit the correct
energy dependence. Indeed at the highest energy eq.(\ref{micro-expan}) strongly
overshoots the expanding simulation.
\begin{figure}[b!]
\setlength{\abovecaptionskip}{-10pt}
\begin{center}
\includegraphics[angle=-90, width=1.1\columnwidth, trim=2.9cm 0cm 1.5cm 0cm,clip=true]{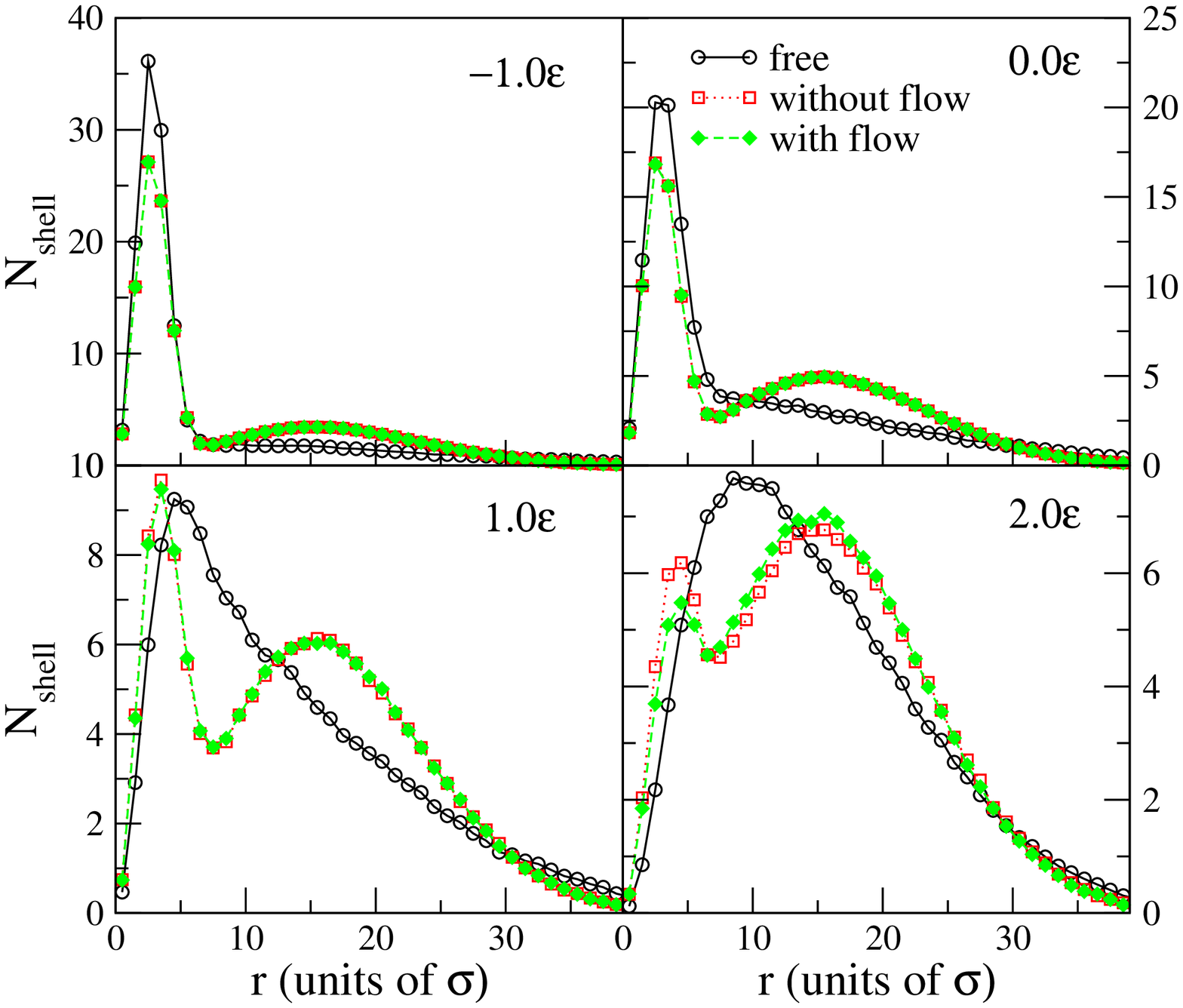}
\includegraphics[angle=-90, width=1.1\columnwidth, trim=2.9cm 0cm 1.5cm 0cm,clip=true]{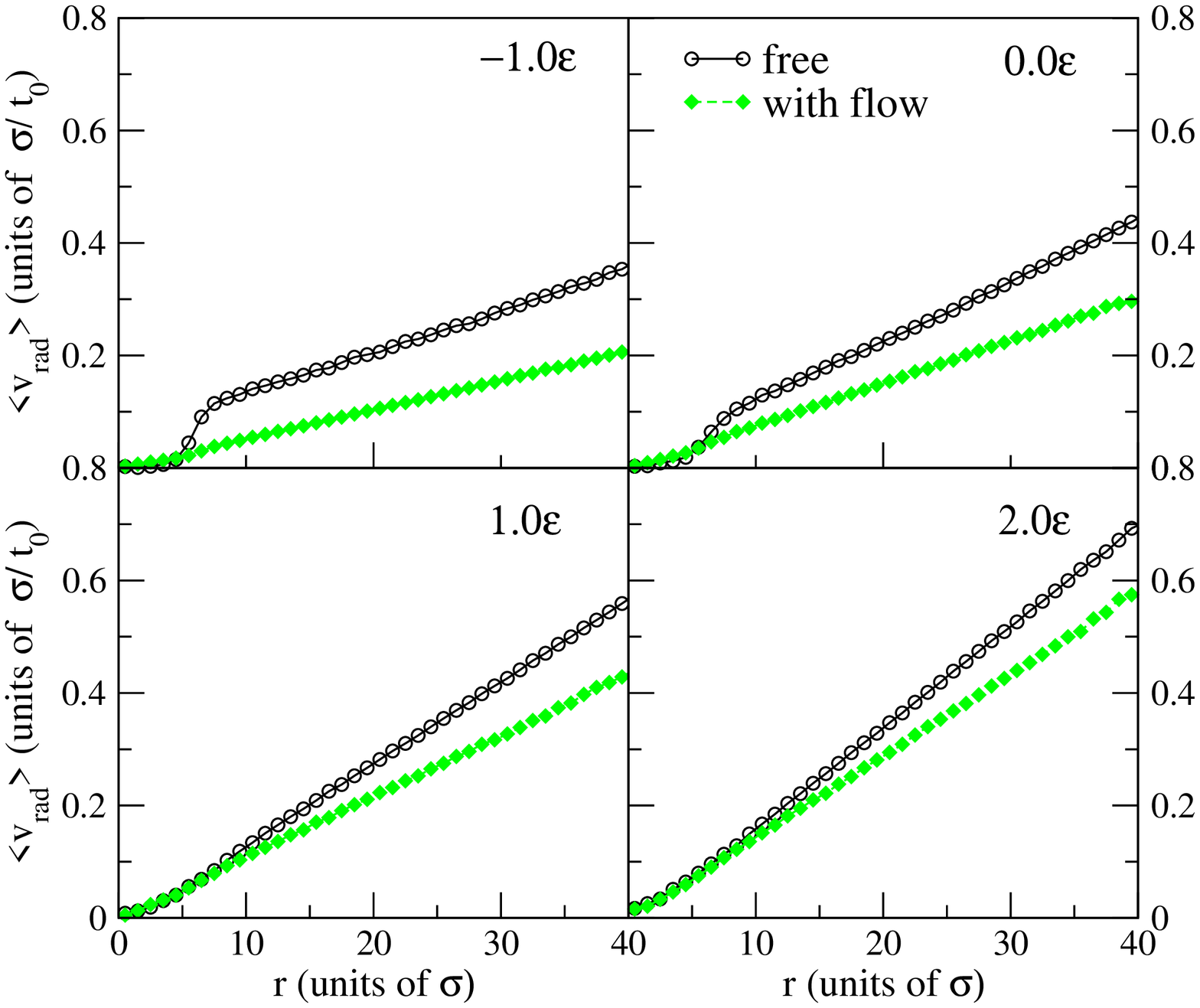}
\end{center}
\caption{\label{fig:fig5}
(Color online) Upper part: average number of particles located in a spherical shell of radius $r$ as 
a function of $r$, calculated from the center of mass of the system. Lower part: average velocity 
in the radial direction in the same spherical shells as the upper part.
The 4 energies $(-1.0,0.0,1.0$ and $2.0\epsilon)$ and the three models (free expansion, equilibrium 
in the local rest frame with and without flow fluctuations) are displayed.}
\end{figure}

To understand the origin of this deviation, we plot in Figure 5 the matter density 
and radial velocity profiles
associated to the four different energies and the three models shown in Figure 4.
The statistical simulations always present a well pronounced density peak close to the center of 
mass of the system, corresponding to a single
drop of decreasing size. The rest of the matter concentrates on a second density 
peak around $r\approx15\sigma$, where most of the clusters are located.

The spatial distribution of matter for the freely expanding system is close to this 
picture at the lowest energy under analysis. However the presence of a heavy drop 
close to the center of mass modifies the shape of the radial flow, which is very far 
away from the self-similar assumption of eq.(\ref{micro-expan}).
The Hubblian flow in eq.(\ref{micro-expan}) comes from the ideal-gas equations of
motion eq.(\ref{eqdyn}). We can see from Figure 5 that the inter-particle $LJ$ 
interaction cannot be neglected in the dense part of the system, 
which leads to an important deviation from self-similarity.
We expect that a better description of the expansion dynamics would 
be obtained considering for the flow dynamics in eq.(\ref{micro-expan}) 
clusters instead of particles degrees of freedom, as in the standard Fisher
picture of condensation \cite{Fisher}. 

At higher energies, the distribution of the free expanding system is completely different
from the equilibrium picture. The central drop progressively disappears leading
to a smoother matter distribution along the radial direction. From previous studies \cite{Dorso},
we know that the resulting large bump corresponds to a high multiplicity of 
approximately equal sized clusters. The presence of clusterized matter leads
to a deviation from the Hubblian flow even at the highest energy.
%
%
\begin{figure}[t]
\setlength{\abovecaptionskip}{0pt}
\includegraphics[angle=0, width=0.9\columnwidth, trim=0cm 0cm 0cm 0cm,clip=true]{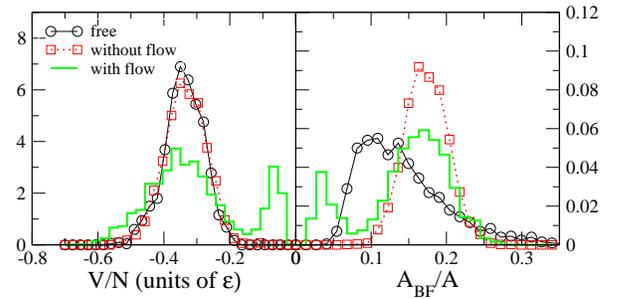}
\caption{\label{fig:fig6}
(Color online) Distributions of potential energy (left panel) and of the size of the biggest MST 
fragment (right panel) for the highest energy considered $(2.0\epsilon)$ and the 
three models (free expansion, equilibrium in the local rest frame with and without flow fluctuations). } 
\end{figure}

More information on the specific configurations accessed by the flow dynamics
can be obtained from Figure 6. This figure displays
the distribution of the potential energy and of the size of the largest 
fragment recognized through the MST algorithm \cite{Dorso}.
It confirms that the freely expanding system explores
configurations which are very different from the equilibrium case.
As already observed in Figure 2, the global energy sharing between potential
and kinetic energy of the free expanding system is consistent with equilibrium, 
and consequently the potential energy distributions are close.
However the associate partitions are very different and tend to be more fragmented
in the free expanding case, as shown by the fact that the largest cluster size 
distribution is broader and peaked at a lower value.
At the energy shown in the figure, the system is close to the liquid-gas
phase transition. If flow fluctuations are allowed, as in eq.(\ref{micro-expan}),
some gas partitions can then be explored, leading to the bimodal distributions
observed in Figure 6. 
The fact that a statistical treatment of flow can lead to bimodality was recently 
observed in ref.\cite{matias}. The presence of bimodal distributions explain the 
severe overestimation of the free expanding system fluctuations shown in Figure 4.
Indeed, for these distributions to be accessed during the actual time evolution of 
the expansion a potential energy barrier has to be overcome, 
and this does not seem to occur easily for the free expanding system. 
As it was already shown in ref.\cite{Chernomoretz}, flow acts as a heat sink 
preventing the exploration of gas configurations. The largest cluster distribution shows
that most partitions of the freely expanding system present an intermediate 
degree of fragmentation between liquid and gas. Such partitions are metastable or
unstable at equilibrium, but can be accessed in the free expansion due to the short
time scale of the dynamics.
%
%
%
\section{Conclusions}

To conclude, in this paper we have compared the diabatic expansion dynamics 
of a Lennard-Jones system, initially confined in a harmonic oscillator and 
subsequently expanding freely in the vacuum, with a statistical ansatz 
in the hypothesis of a purely Hubblian flow. This hypothesis is exact in the 
limiting case of a non-interacting system or a Boltzmann dynamics.
For our strongly interacting system, the presence of finite range two body interactions
is known \cite{annals} to modify the Hubblian approximation introducing 
non-self-similar flow components. In a future work, it will be very interesting
to explore the adequacy of a more sophisticated statistical ansatz
including non-self-similar flows, to reproduce the dynamics of the expansion.
In the present paper, we have limited ourselves to the self-similar approximation
which reveals to be accurate enough to reasonably describe the mean value of global
one-body observables at all times.
As soon as more sophisticated observables are examined, discrepancies arise.
In the self-similar statistical ansatz, flow does not modify the partitions
in configuration space, but it acts as a heat bath allowing important energy 
fluctuations and the exploration of the unbound gas phase. 
In the diabatic dynamics, such configurations are never reached and metastable
highly clusterized partitions dominate.
It is interesting to observe that qualitatively similar behaviors
have recently been observed in an analysis of nuclear multifragmentation 
data \cite{nicolas}, by means of a detailed comparison between the fragmentation 
of central and peripheral collisions.
  
%
%
%
\bigskip
\section{Acknowledgements}

Partial support from the University of Buenos Aires via Grant X360 is acknowledged. 
MJI also acknowledges the kind hospitality of the LPC, where the core of the work was 
developed, as well as financial support from the LPC, the University of Buenos Aires, and 
Fundacion Antorchas.

%

%
%
%
\end{document}